\documentclass[a4paper]{article}

\usepackage[dvips]{graphicx}
\usepackage{amsmath,amssymb}
\usepackage{ulem}
\usepackage{color}\usepackage[many]{tcolorbox}
\usepackage[round]{natbib}
\makeatletter
\renewcommand\@biblabel[1]{}
\makeatother

\newtcolorbox{boxA}{
    boxrule = 1.5pt,
    colframe = black 
}

\title{Integrated information theory: the good, the bad and the misunderstood}

\author{Adam B.~Barrett$^1$\footnote{adam.barrett@sussex.ac.uk}, Borjan Milinkovic$^{1,2,3}$, Pedro A.~M.~Mediano$^4$, \\
Fernando E.~Rosas$^{1,5,6}$, Daniel Bor$^7$, Lionel Barnett$^1$
and Anil K.~Seth$^{1,8}$\\\\
\small
$^1$\textit{Sussex Centre for Consciousness Science and Department of Informatics,}\\
\small \textit{University of Sussex, Brighton, UK}\\
\small$^2$\textit{Institute of Neurosciences (NeuroPSI), CNRS, Gif-sur-Yvette, Paris, France} \\
\small$^3$\textit{Paris Brain Institute (ICM), Inserm, Hôpital de la Pitié-Salpêtrière,}\\ \small\textit{Paris, France}\\
\small$^4$\textit{Department of Computing, Imperial College London, London, UK}\\
\small$^5$\textit{Center for Psychedelic Research and Centre for Complexity Science,} \\\small\textit{Department of Brain Science, Imperial College London, London, UK}\\
\small$^6$\textit{Centre for Eudaimonia and Human Flourishing, University of Oxford, Oxford, UK}\\
\small$^7$\textit{Department of Psychology, Queen Mary University of London, London, UK}\\
\small$^8$\textit{Program on Brain, Mind, and Consciousness,} \\ \small\textit{Canadian Institute for Advanced Research, Toronto, Canada}
}
\date{}

\begin{document}

\maketitle

\begin{abstract}
\noindent The integrated information theory of consciousness (IIT) is uniquely ambitious in proposing a mathematical formula, derived from apparently fundamental properties of conscious experience, to describe the quantity and quality of consciousness for any physical system that possesses it. IIT has generated considerable debate, which has engendered some misunderstandings and misrepresentations. Here we address and hope to remedy this. We begin by concisely summarising the essentials of IIT. Given IIT is supposed to apply universally, we do this with reference to an arbitrary patch of matter, as opposed to the usual system of discrete computational units. Then, after briefly summarising IIT's theoretical and empirical achievements, we focus on five points which we consider especially important for driving forward new theory and increasing understanding. First, a high value of the measure $\Phi$ is not synonymous with `more consciousness'. We describe how $\Phi$ might be replaced with a suite of quantities to obtain a multi-dimensional characterisation of states of consciousness. Second, we describe with nuance the distinct flavour of panpsychism implied by IIT -- whereby space (and time) are tiled with substrates of (proto-) consciousness -- and find this is not problematic for the theory. Third, $\Phi$ is not well-defined for real physical systems, and has not been computed on any real physical system. Fourth, so far only \textit{proxies} for IIT measures have been computed, and not \textit{approximations}. Fifth, for IIT to fit with current successful theories in fundamental physics, a reformulation in terms of continuous fields would be needed. 
\end{abstract}


\section{Introduction}

\noindent Integrated information theory (IIT; \citealt{Albantakis2023,Tononi2016}) continues to garner substantial interest and attract controversy for its claim to comprehensively explain the physical basis of consciousness  \citep{Hakwan,Tononi2025,Gomez-Marin2025}. Several technical and philosophical challenges to the theory have been identified \citep{Bayne2018, Barrett2019, Morch2019, Mediano2022,Cea2023}, yet there still remains a sense among many researchers that its approach holds promise. Unfortunately, debates about IIT are frequently characterised by what we take to be misunderstandings of the theory, holding back progress. In this article, our aim is to remedy some of the most common and significant of these misunderstandings. By doing so, we hope to increase understanding of both the achievements and potential of IIT, as well as some of the theoretical challenges pertaining to its formalism, in the hope of driving consciousness research forward.

The article begins with a conceptual overview of the essential components of IIT. Taking account of the premise that IIT applies universally, this is done with reference to an arbitrary patch of matter, as opposed to the usual system of discrete computational units. Then follows some description of IIT's most important achievements, focusing on empirical evidence inspired by IIT and compatible with it, and on the ability of the theory to link specific neural structural and dynamical features to specific phenomenal properties. We then dispel two common misconceptions: \textit{(i)} that high $\Phi$ is exactly synonymous with  `more consciousness'; and \textit{(ii)} that the form of panpsychism implied by IIT is distinctively problematic. Next, three challenges for IIT are addressed: \textit{(a)} the IIT algorithm, and $\Phi$ itself, are not sufficiently well-defined; \textit{(b)} so far only \textit{proxies} for IIT measures (such as $\Phi$) have been computed, and not approximations; and \textit{(c)} for IIT to fit with current successful theories in fundamental physics, a reformulation in terms of continuous fields would be needed. The article concludes with a brief Discussion, which includes a list of recommendations (Box 2).

\section{The essentials of IIT}

IIT has a rich history. An appropriate starting point is the dynamic core hypothesis proposed by Giulio Tononi and Gerald Edelman in the late 1990s \citep{Tononi1998}. This hypothesis proposed that an explanatory link could be drawn between properties characteristic of all conscious experiences, and properties of neural dynamics, when formalised mathematically. Specifically, the hypothesis proposed that every conscious experience is both integrated (experienced as a whole) and differentiated (every experience is one among a vast repertoire of possible experiences), and that this combination of properties should be reflected in the \textit{dynamical complexity} of its neural basis, where dynamical complexity is a mathematical measure that maximises co-existing integration and differentiation in the dynamics of a network. The dynamic core hypothesis commits to an explanatory relation \citep{Seth2009}, between neural processes and conscious experience, but does not specify anything about a sufficient basis for consciousness. The introduction of a necessary and sufficient basis for consciousness came later, with the first emergence of IIT itself, starting in 2004 \citep{Tononi2004}. 

The current version of IIT, IIT 4.0, takes as its starting point six properties of conscious experiences that it considers fundamental, referring to them as `axioms' \citep{Albantakis2023} - see Box 1. It then derives from these a universal algorithm for describing the quality and quantity of consciousness for any physical system \citep{Albantakis2023}. The key outputs of the algorithm are: system integrated information $\varphi_s$; cause-effect structure $C$; and structure integrated information $\Phi$. The formulae for these are given in full in \citep{Albantakis2023}, and a concise summary of how they are derived is given in \citep{Mediano2022}. We emphasise that the IIT algorithm is provided as a means by which researchers can calculate or approximate quantities related to the conscious experience associated with a physical system. The theory does not imply that consciousness itself is algorithmic. In fact, IIT is a canonically non-computational, non-algorithmic theory \citep{findlay2024dissociating, Albantakis2023,Tononi2025,Gomez-Marin2025}. The following paragraphs outline the key conceptual components to the IIT algorithm.

\begin{boxA} \label{box:axioms}
\subsubsection*{Box 1. The fundamental properties of conscious experience according to IIT.}
IIT is constructed from considering as fundamental the following set of six properties of conscious experience, which are referred to as `axioms'. These have sometimes been misinterpreted, so we write them here together with some points of clarification, that draw from the IIT wiki (at \texttt{https://www.iit.wiki/}). The formal stated axioms are in italics, and the points of clarification in normal font.  
\begin{itemize}
\item \textbf{Existence.} \textit{Experience exists:} For each experience, \textit{there is something.}
\item \textbf{Intrinsicality.} \textit{Experience is intrinsic: it exists for itself.}
\item \textbf{Information.} \textit{Experience is specific: it is this one.} Given the occurrence of one experience, this rules out the occurrence of other potential experiences that could a priori possibly have been entailed by the physical substrate.
\item \textbf{Integration.} \textit{Experience is unitary: it is a whole, irreducible to separate experiences.} Trying to conceive of an experience that were not unitary leads to conceiving of two separate experiences, each of which is unitary.
\item \textbf{Exclusion.} \textit{Experience is definite: it is this whole.} Given an experience which does exist, there do not exist simultaneous sets of experiences with more or less content. Note that, unlike the other axioms, this is not obvious from introspection, yet it constitutes a reasonable assumption.
\item \textbf{Composition.} \textit{Experience is structured: it is composed of distinctions} (distinct phenomenal components) \textit{and the relations that bind them together, yielding a phenomenal structure that feels the way it feels.} For example, within a single experience, one may distinguish a piano, a blue colour, a book, several spatial locations, sounds, and various emotions. Note that IIT allows for the possibility of experiences with just a single phenomenal component, but it doesn't allow experiences with no components at all.
\end{itemize}
\end{boxA}

\vspace{0.3cm}

The system integrated information $\varphi_s$ is used to identify precisely where within a system consciousness arises. For there to be consciousness associated with a given patch $P$ of physical matter, the system integrated information of the patch $\varphi_s(P)$ must be greater than that of all other patches that overlap with $P$, see Fig.~1(a)-(d). Note we use the word \textit{patch} here to emphasise how the IIT algorithm can in theory be applied to the constituents of any patch of space; we use the word \textit{system} in place of patch when the emphasis is on applying IIT to a specific object of study.

Once a patch of physical matter is associated with consciousness, following the above specification, the qualities of the corresponding conscious experience (conscious contents) are described by the cause-effect structure $C$ of the patch, while the overall amount, or level, of consciousness is given by the structure integrated information $\Phi$.

Importantly, the output of the algorithm will depend on three `grainings' \citep{Marshall2024}: \textit{(i)} how the contents of the patch are spatially specified, i.e., the graining of the patch into a set of discrete components [Fig.~1(e, f)]; \textit{(ii)} the size of the time difference between the past, present and future states considered, and; \textit{(iii)} the definition of the states of the components, e.g., binary presence or absence of an action potential, or number of action potentials in a population of neurons within a defined time-interval. The grainings that determine the presence and qualities of consciousness are those that maximise the integrated information measures $\varphi_s$ and $\Phi$.

\begin{figure}
\begin{center}
\includegraphics[width=0.96\textwidth]{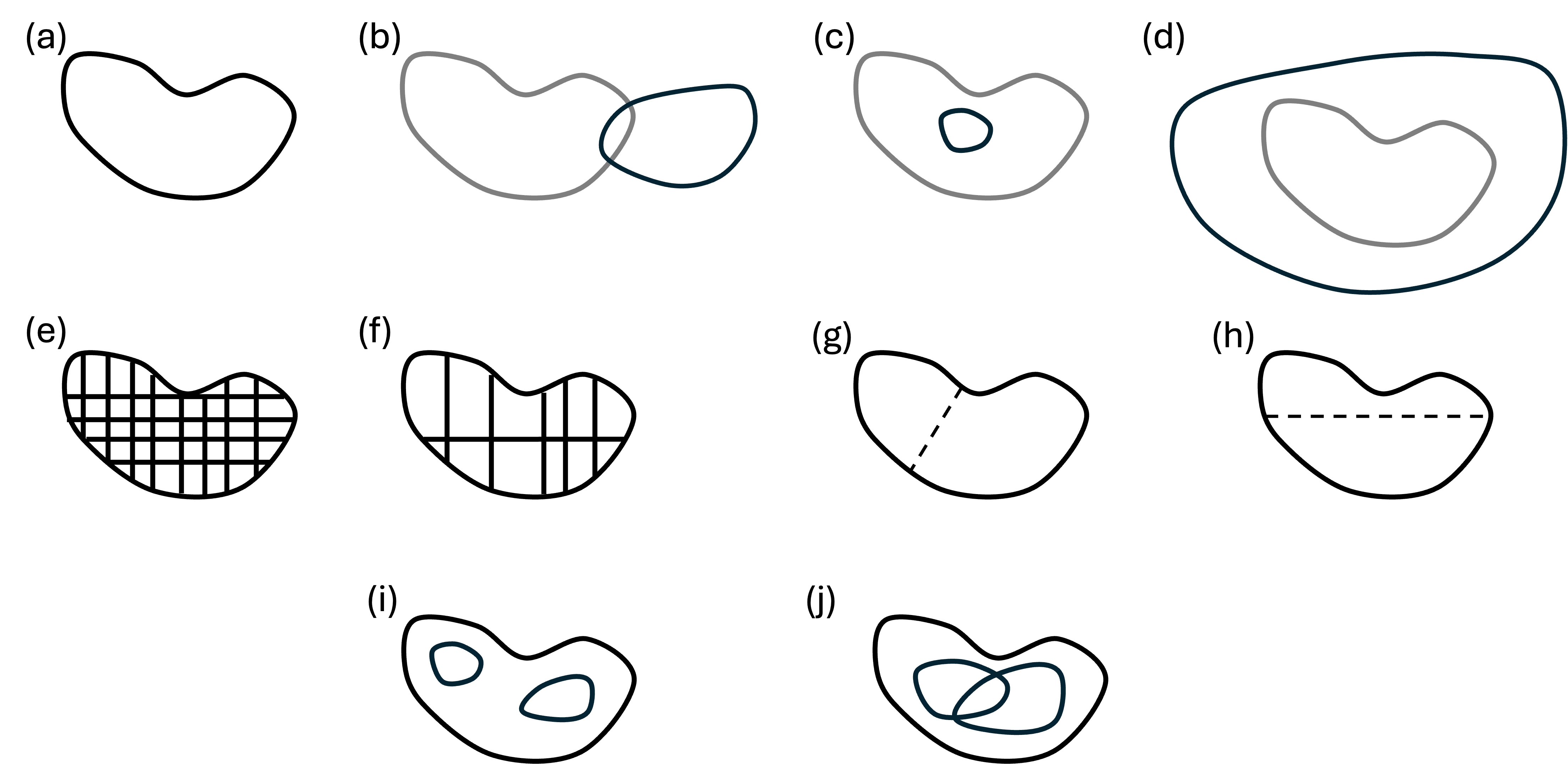} \label{fig:IIT_on_patch}
\end{center}
\caption{Schematic for conceptualising the IIT algorithm being applied to the physical matter within an arbitrary patch of space. (a) An example patch $P$ to which the IIT algorithm is applied. (This patch appears 2-dimensional here for ease of illustration; it should however be thought of as 3-dimensional.) Patch $P$ is a substrate of consciousness if and only if its system integrated information $\varphi_s(P)$ is greater than that of all other patches that overlap with $P$, such as those shown in panels (b)-(d). Whether or not a patch is a substrate of consciousness varies over time. The outputs of the IIT algorithm depend on how the contents of the patch are `grained' into a set of discrete `components'. Components need not all be the same size or shape; two example grainings of $P$ are shown in panels (e) and (f). The graining that determines whether $P$ is a substrate of consciousness is that which maximises $\varphi_s$. System integrated information $\varphi_s$ derives from information that the current state of the whole patch contains about states of the whole patch one time-step into the past and future, that cannot be attributed to parts of the patch when the patch is partitioned. Panels (g) and (h) show two example partitions of $P$. All partitions into an arbitrary number of parts are considered (not just bipartitions as shown here). The cause-effect structure $C$, which determines the contents of consciousness entailed by patches that are substrates of consciousness, derives from the detailed informational relationships between past, present and future states of all pairs of parts of the patch. Overlapping and non-overlapping parts are considered, as illustrated in panels (i) and (j).}
\end{figure}

All of the key quantities in the algorithm derive from informational relationships between past, present and future states of the patch and parts of the patch (by \textit{part} we mean any sub-patch; these generally contain multiple components). Specifically, they derive from the probability of occurrence of each past and future state of some part of the patch given the current state of some part of the patch. Crucially, these are calculated assuming that all states are a priori equally likely (in technical language, a `maximum entropy' prior). The system integrated information $\varphi_s$ derives from the information that the current state of the whole patch contains about states of the whole patch one time-step into the past and future, that cannot be attributed to parts of the patch [Fig.~1(g, h)]. The cause-effect structure $C$ derives from the detailed informational relationships between past, present and future states of all pairs of parts [Fig.~1(i, j)]. Its contents include sets of parts, their overlaps, and a collection of information-theoretic quantities. Geometrical and topological properties of the cause-effect structure correspond to specific aspects of the contents of consciousness. (For understanding IIT, it is important to remember that the cause-effect structure $C$ is distinct from physical causal structure, the latter usually being defined as a set of physical causal interactions between components within a system.) Finally, the structure integrated information $\Phi$ sums up the integrated information associated with each set of parts in the cause-effect structure.


The above quantities are computed based on the current state of the patch and the causal mechanisms that exist within the patch at the current time, as determined by transition probability matrices. From one moment to the next, as dynamics occur, all three of the key quantities will change and evolve. Within some physical system, the precise patches entailing consciousness will thus typically be constantly changing (according to whichever non-overlapping patches have maximal $\varphi_s$), and the quantity and qualities of consciousness will also fluctuate with a system's dynamics.

\section{Ambitions and achievements of IIT}

IIT is unique among theories in its approach to consciousness. No other theory tackles the question of identifying the physical substrate of consciousness by starting from the fundamentals of phenomenology, see \citet{Seth2022} and \citet{Kuhn2024} for comparative reviews. In other words, IIT addresses the `hard problem' backwards \citep{Chalmers1995}; by proceeding from phenomenological axioms to mechanisms, as opposed to trying to go from physical mechanisms to consciousness. Given the intransigence of consciousness to standard approaches, this creative take on the hard problem is worth appreciating. In addition, IIT is uniquely comprehensive as  a theory of consciousness, providing accounts of: \textit{(i)} what is and isn’t conscious; \textit{(ii)} the existence of distinct levels of consciousness; and \textit{(iii)} specific conscious contents. In this section we briefly discuss some notable accomplishments of IIT: first with respect to obtaining empirical validation; and second at linking neural structures to qualia.

\subsection{Empirical evidence consistent with IIT}

IIT has inspired important empirical findings about the neurobiological basis of consciousness. Perhaps the most notable example is the series of experiments that examine the electroencephalographic (EEG) response to transcranial magnetic stimulation (TMS; \citealt{Massimini2009}). These studies found that when participants are unconscious, whether in anaesthesia \citep{Ferrarelli2010}, sleep \citep{Massimini2005}, or disorders of consciousness \citep{Casarotto2016}, the brain's response to TMS is stereotypical across electrodes and remains local to the site of stimulation (at standard stimulus intensities). By contrast, for conscious participants, the brain response to TMS is more diverse across electrodes and spreads across larger regions of cortex. In these experiments, this effect was quantified using the perturbational complexity index (PCI), which quantifies the signal diversity of the EEG response to the TMS \citep{Casali2013}. PCI is reliably lower in deep sleep and anaesthesia states than in wakeful rest, as predicted by IIT. Moreover, PCI has found clinical application in the assessment of conscious level in brain-injured patients suffering disorders of consciousness \citep{Casarotto2016,Re2021}. A detailed summary of these experiments and their findings is given in \citet{Sarasso2021}. 

The PCI results are encouraging for the IIT enterprise, carry significant practical implications, and shed light on the neurodynamic underpinning of global states of consciousness. In particular, they give validation to the information and integration axioms, and the possibility to map these to measurable quantities derived from physical phenomena. However, they do not provide a strong test of all the specific and distinctive claims of IIT, as expressed by the IIT algorithm; these results are equally compatible with alternative weaker versions of IIT, as well as, arguably, with other theories of consciousness such as global neuronal workspace theory \citep{Farisco2023,Mudrik2025}.

It has thus recently been proposed to make a distinction between `weak' and `strong' flavours of IIT \citep{Mediano2022} -- with strong IIT \textit{strictly identifying} conscious experiences universally with the theoretical quantities of IIT (and the IIT algorithm providing the in-principle means to measure these quantities), and weak IIT focusing on \textit{explanatory correlates} of consciousness \citep{Seth2009} that hold only in certain specific experimental scenarios - echoing the origins of IIT in the dynamic core hypothesis \citep{Tononi1998}. The strength of the weak approach is that it provides a theoretically more conservative account for the PCI results; and one where the metaphysical claims of strong IIT are not needed. Nevertheless, it is hard to imagine the PCI measure ever having been developed without the prior development of at least some version of IIT.






\subsection{IIT can link neural structures to qualia}
It is a notable achievement of IIT to provide an explanation of how: \textit{(i)} the spatial aspect of visual experience may arise from a network of neurons with a grid-like connectivity structure \citep{Haun2019} [Fig.~2(a)]; and \textit{(ii)} the feeling of temporal flow may arise from a network of neurons arranged in a directed chain \citep{Comolatti2024} [Fig.~2(d)]. In addition, ongoing research is attempting to link certain architectures to narrow qualia (e.g.~colour) and the presence of conceptual hierarchy. None of the other major theories of consciousness have developed accounts of these widespread properties of conscious experiences.

\begin{figure}
\begin{center}
\includegraphics[width=0.96\textwidth]{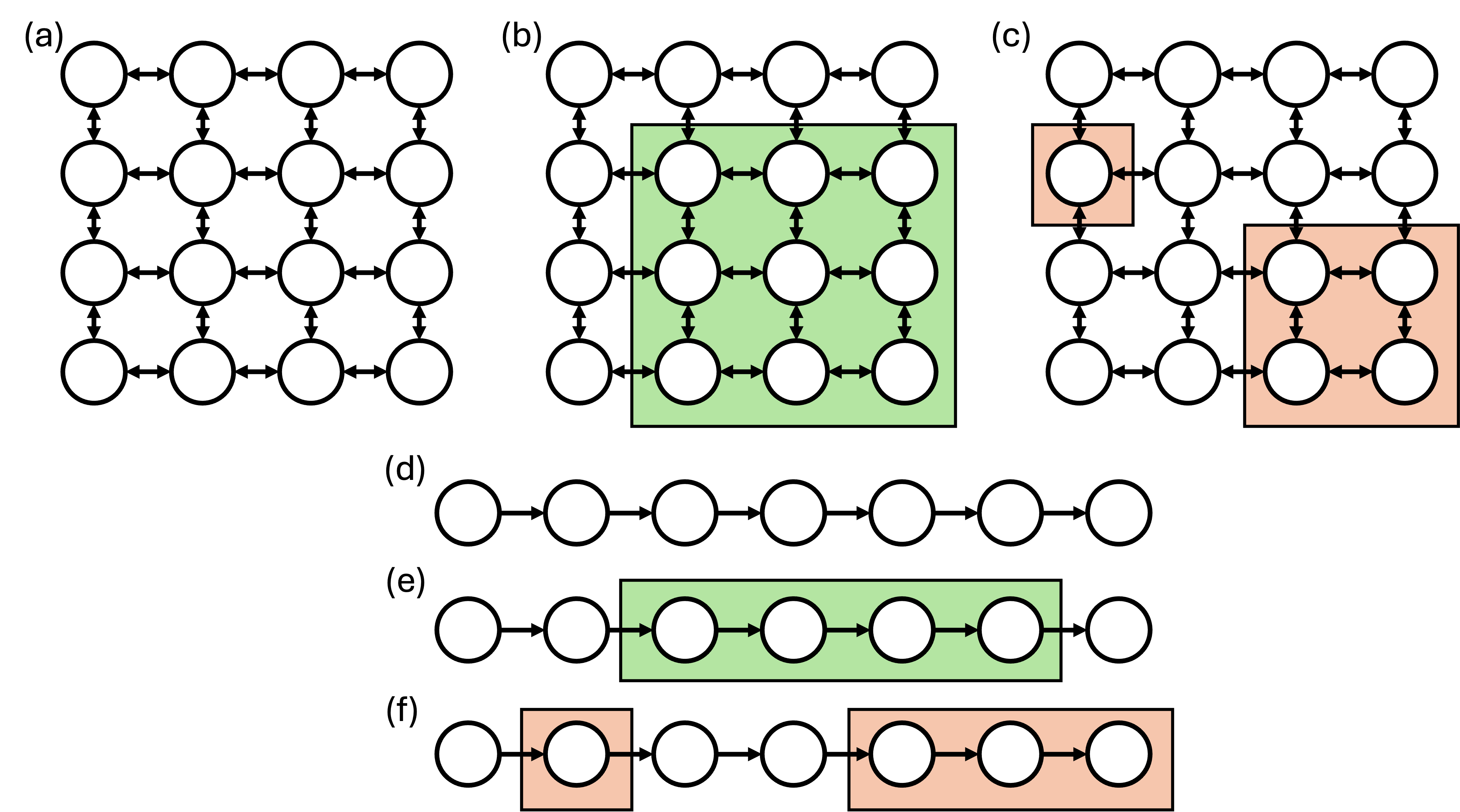} \label{fig:qualia}
\end{center}
\caption{Neural architectures associated with experiences of space and time. In this figure, circles represent indivisible system components that update their states in discrete time. Arrows represent connections between components - the state of one component at the next time-point depends on the current states of the components to which it is connected with an incoming arrow. (a) For such a system with components arranged as a non-directed grid, an experience of space is entailed. (b) Contiguous parts of such a system, such as that shaded in green, cannot be partitioned in two without there being a loss of information about what the next state will be, as some connections would necessarily have to be cut. Thus contiguous parts contribute to the cause-effect structure. (c) Non-contiguous parts, such as that shaded in orange, can be partitioned without there being a loss of information about what the next state will be; they can be partitioned according to where they are non-contiguous (here the single component to the left and the set of 4 components to the right). Thus non-contiguous parts do not contribute to the cause-effect structure (as a whole - each of their contiguous groups of components do, independently). (d) For such a system with components arranged as a directed chain, an experience of temporal flow is entailed. (e) Contiguous parts of such a system, such as that shaded in green, cannot be partitioned in two without there being a loss of information about what the next state will be, and thus they contribute to the cause-effect structure. (f) Non-contiguous parts, such as that shaded in orange, can be partitioned without there being a loss of information about what the next state will be (here the single component to the left and the set of 3 components to the right), and thus do not contribute to the cause-effect structure (as a whole - each of their contiguous groups of components do, independently).}
\end{figure}

It is worth noting that this accomplishment does not require the ability to measure $\Phi$ itself (which, as we discuss later, is problematic). This is because, for the category of toy models utilised, the cause-effect structure can be specified without computing $\Phi$. These systems consist of components that update their states synchronously in discrete time, according to rules specified by connections between components. The cause-effect structure for such systems is determined simply by which sets of components cannot be partitioned into two without incurring some information loss from the cutting of direct connections between components. Specifically, for the architectures considered, contiguous sets of components cannot be partitioned into two without cutting direct connections between components, whilst non-contiguous sets of components can (precisely where they are non-contiguous), see Fig.~2(b, c, e, f). 




For such a system with the architecture of a non-directed two-dimensional grid, IIT associates each contiguous set of components to a region in the spatial canvas of a visual experience [Fig.~2(a-c)]. This is done by recognising that the overlaps between all the pairs of contiguous sets of components has exactly the same topology (geometrical structure) as the overlaps between the various spots one can distinguish on the spatial canvas of one's visual experience \citep{Haun2019}. For a directed chain, there is a directionality to the cause-effect structure, arising from the asymmetry between the set of components that influence, and are influenced by, each set of components [Fig.~2(d-f)]. This can then be associated with the phenomenology of temporal flow \citep{Comolatti2024}. 

\section{Common misunderstandings of IIT}

We now discuss two common misunderstandings of IIT that, in our experience, have often led to inaccurate critiques of the theory. 

\subsection{High $\Phi$ cannot be equated with more consciousness.} \label{sec:highphi}
The first misunderstanding arises from thinking that subscribing to IIT necessarily implies supporting the notion of a one-dimensional level of consciousness, with $\Phi$ being the measure for this. While IIT does state that $\Phi$ corresponds to the quantity of consciousness \citep{Albantakis2023}, there is scope to subscribe to IIT with a multi-dimensional view of global states of consciousness, such as that of \citet{Bayne2016}. (Following \citet{Bayne2016} we take the global state of consciousness to characterise the overall (un)conscious condition of a system, or organism. Examples of global states of consciousness are NREM sleep, wakeful rest and general anaesthesia.) $\Phi$ is a dynamical variable, thus one `reading' of $\Phi$ constitutes a measure of the quantity of consciousness entailed by a system at just one moment in time. However, in theory one could use multiple samples of $\Phi$ over a period of time, as well as any of the quantities in the cause-effect structures at multiple points in time to construct a multi-dimensional characterisation of the global state of consciousness of a system. For example, one could take: \textit{(i)} the mean $\Phi$ over a period of time as a measure of mean `intensity' of consciousness (indeed, this would be one way of bringing precision to this concept); \textit{(ii)} variances over time of quantities within the cause-effect structure to quantify how much the contents of consciousness are varying over time; \textit{(iii)} some measure of (mean) complexity of the geometry and topology of the cause-effect structures (complexity arises there from inhomogeneity in the connections between system components) to characterise the complexity of the phenomenology. 

If one takes such a multi-dimensional stance towards describing global states of consciousness with IIT, then the often-cited `expander grids'  no longer cause the severe problems for the theory that were identified by \citet{Aaronson}. Expander grids are two-dimensional grids of logic gates with each logic gate connected to each of its nearest neighbours [as in Fig.~2(a)]. An expander grid on which all of the logic gates are inactive has an unboundedly high $\Phi$ as the number of logic gates increases. This poses a plausibility challenge for IIT if high $\Phi$ is straightforwardly taken to mean `lots of consciousness'. However, taking the alternative multi-dimensional view leads to the less problematic conclusion that there is in fact minimal consciousness. Here is why.

First, the contents of consciousness entailed by an inactive expander grid never evolve. For any changes to occur in the contents of consciousness, the physical substrate needs to be embedded in a dynamical process, i.e., the system must transition into a new state. A grid of inactive logic gates will, if left on its own, continue to exist in an inactive state, and thus the entailed contents of consciousness will never evolve. IIT would posit that the grid has a conscious experience with a high $\Phi$ when it is first assembled, but then it would not have any further conscious experiences as long as it remains inactive. 

Second, the simple grid structure would entail phenomenology of an empty spatial field with nothing in it, and no other component to the experience - not even time \citep{Haun2019}. Thus, on IIT, in spite of having a high $\Phi$, inactive expander grids entail only a phenomenologically minimal form of conscious experience, no matter how large the grid becomes. 

Thus, according to the multidimensional three-component characterisation of consciousness proposed above, only mean $\Phi$ would be large, while the variances of cause-effect structure quantities would be zero, and complexity of the cause-effect structure would be minimal (however this is measured). An expander grid having some quantity of consciousness may still be considered a problem, or at least counterintuitive, but it is a problem (or counterintuitive challenge) of a different and lesser order than the idea that inactive grids have  unboundedly high quantities of consciousness.
 


Debates about expander grids arose out of the distinctive feature of IIT that inactive components in a system can contribute to conscious experience. While much has been written about this feature \citep{Bartlett2022}, it has been less frequently appreciated that a large number of components must exhibit dynamic behaviour to entail a global state of consciousness in which there is complex phenomenology. It is not enough to achieve high $\Phi$ once, by simply existing in one single static physical state. This point has not been much discussed in the IIT literature, yet it is naturally implied by the theory. 

Taking a multi-dimensional approach to global states of consciousness does help understand IIT and its implications. But it also reveals another potential weakness. It is hard to see how IIT could capture all dimensions that might be considered as relevant explanatory targets, in particular, dimensions related to metacognition, sense of self, or capacity for suffering -- all of which are prominent in human consciousness.

We summarise this subsection as follows. Global states of consciousness can only be properly characterised in a multi-dimensional framework \citep{Bayne2016}. It is not well-defined what aspect of phenomenology $\Phi$ itself captures. There are other potentially useful dimensions that could be derived from cause-effect structures, providing opportunities for future research. However, it is unlikely that these will be able to capture all dimensions of interest; and this challenges the notion of IIT being a fully comprehensive theory of consciousness. Expander grids with high $\Phi$ only challenge the utility of $\Phi$ itself, and not the entirety of IIT.

\subsection{The panpsychism of IIT is distinct from that of other theories, and is not a problematic metaphysical position} \label{sec:panpsychism}
IIT has been frequently criticised for implying panpsychism \citep{Hakwan}. Panpsychism is, broadly, the view that consciousness is a fundamental property of matter, and that it exists essentially everywhere. This might sound implausible, but most, if not all metaphysical positions for explaining consciousness seem rather implausible when held up to the light \citep{Kuhn2024}. For example, the standard materialist view requires the emergence/entailment/identification of consciousness from (or with) physical matter that in itself is unconscious. There is no prominent theory that provides any potential explanation of the fundamental mechanism by which this happens - as reflected by the persistence of the `explanatory gap' \citep{Levine1983} and the `hard problem' of consciousness  \citep{Chalmers1995}. 

\begin{figure}
\begin{center}
\includegraphics[width=0.96\textwidth]{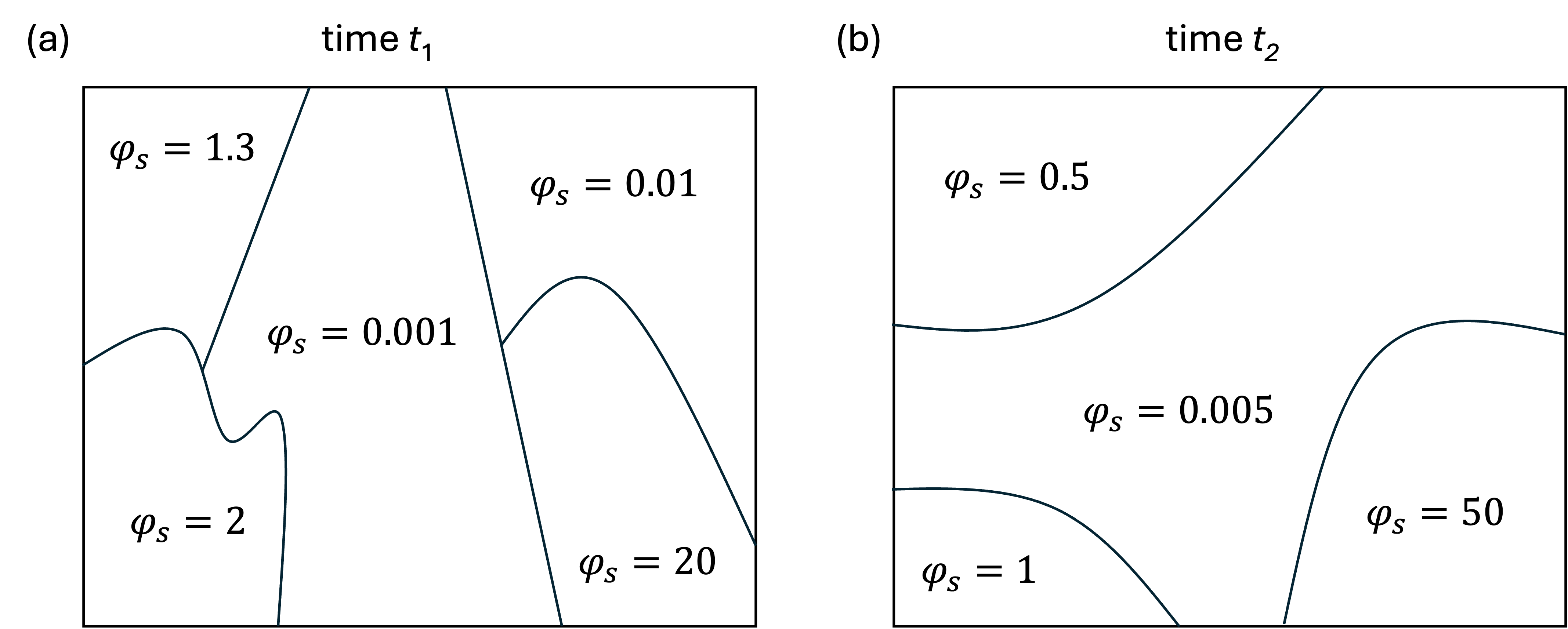} \label{fig:panpsychism}
\end{center}
\caption{The panpsychism of IIT. IIT implies a tiling of space and time with substrates of consciousness (with no gaps). As matter moves and interacts, the boundaries of distinct substrates shift around. Here are two schematics of a region of space showing two hypothetical tilings at two moments in time. Mostly there are only micro forms of consciousness, or `ontological dust'. However, when matter is arranged in a very complex structure, a cause-effect structure with substantial conscious content can emerge.}
\end{figure}

IIT does not assume panpsychism, but it does imply a distinct flavour of it that has not been fully explicated in the literature. Each conscious experience that occurs is associated with a certain patch of space and a certain window of time. By the exclusion axiom, these never overlap for distinct conscious experiences. IIT therefore implies a tiling of space and time with substrates of consciousness (with no gaps), see Fig.~3. Moreover, all physical matter is conscious in the sense that it contributes to precisely one substrate of consciousness at any given time. (Precisely, the patch containing it that maximises $\varphi_s$.) As matter moves and interacts, the boundaries of distinct substrates shift around. Any specific physical system of interest will thus at all times either \textit{(i)} contain one or more (non-overlapping) substrates of consciousness, or \textit{(ii)} be part of a larger physical system that contains one or more (non-overlapping) substrates of consciousness. Throughout the physical universe,  there are mostly only micro forms of consciousness, or `ontological dust' \citep{Tononi2022}, i.e.~cause-effect structures with minimal phenomenal content associated with them. However, when matter is arranged in a very complex structure, a cause-effect structure with substantial conscious content can emerge. 
 
For those uncomfortable with the label \textit{panpsychism}, a possible workaround is to substitute the word `consciousness' for `proto-consciousness' (corresponding to ontological dust) throughout most of IIT. The term `consciousness' can then be reserved for cases where the cause-effect structure surpasses a certain level of complexity, and contains certain contents worthy of being compared, at least minimally, to (for example) a typical healthy adult human waking conscious moment. Introduction of the term `proto-consciousness' could have ethical implications, for instance by allaying concerns surrounding the over-attribution of moral patienthood and rights \citep{Hakwan}.



\section{Problems and misrepresentations of IIT}

\subsection{The IIT algorithm is not well-defined for real physical systems and has not been computed on any real-world physical system} \label{sec:undefined}

\subsubsection{The IIT algorithm is not universally well-defined} \label{sec:undefined2}

Although IIT is uniquely ambitious and creative in its explanatory scope, it is not well-defined for real physical systems \citep{Barrett2019}. The reason for this is that the formulae for $\varphi_s$, $C$ and $\Phi$ require obtaining, \textit{for each possible graining}, the probability distribution for the immediate future state given only the present state of the system; but this is not uniquely specified for grainings under which the state dynamics have memory, i.e., are non-Markovian, such that future states depend on past states in addition to the present state. The IIT algorithm is therefore not well-defined whenever there is at least one non-Markovian graining. Since, in general, some grainings of systems are Markovian and some are non-Markovian, this problem applies broadly. In particular, the IIT algorithm does not produce an output for brains, since brain dynamics are non-Markovian under typical grainings used for modelling, see e.g.~\citet{Fulinski:1998}.

In \citet{Barrett2019} it was discussed how one might attempt to fix this by imposing a uniform distribution on the entire past history, but that this does not always lead to convergence to stable outputs as the length of the past history considered is increased (specifically there is no convergence when dynamics are non-ergodic, such as on a random walk). (There was a suggestion in \citet{Kleiner21} that the problem may be solved by working in addition with entire future trajectories instead of just the immediate future state. The problem remains however, that part of the IIT algorithm requires calculation of evolution of the probabilistic state of one part of the system given no knowledge of the present state of another part of the system. To do this, a probability distribution must be specified for the latter, but the former remains unspecified by this if the dynamics are non-Markovian.)

The problem with non-Markovian dynamics arises from the need to quantify the information that the current state holds about some prior state \citep{Barrett2019}. It might be that the only solution is to reformulate the IIT algorithm to operate solely on the geometrical and topological structure of the instantaneous state of the system/patch in question, and entirely avoid the appearance of past and future states. It could be considered that this would considerably change the fundaments of the theory, in that it would de-emphasise the causal mechanisms in a system \citep{Barrett2019}. However, a complex set of causal mechanisms are needed to generate a system that can exist in multiple complex configurations, thus these would still be fundamental to consciousness on such an updated theory.

\subsubsection{The IIT algorithm has not been computed on any real-world physical system}\label{sec:notcomputed}
The IIT algorithm has not been computed on any realistic model of a physical system. The toy model systems on which the IIT algorithm has been applied are composed of a finite set of binary units that interact with each other and update their states in discrete time, according to a memoryless (Markovian) dynamics. It is implicitly assumed that these components cannot be subdivided into anything smaller. Such systems are not realistic models of any real physical system that one would be interested in applying the IIT algorithm to. In particular, neurons, or any modelled units within neurons, contain considerable physical structure that can be sub-divided many times, all the way down to the individual particle level. The IIT algorithm requires maximising $\varphi_s$ and $\Phi$ over the vast space of all possible grainings of space, time and the states of components, yet an investigation into how one could do this has never been conducted. Moreover, such an investigation could not even begin until a solution is found to the problem of how to treat grainings for which the effective dynamics are non-Markovian.

Another point to note here is that natural choices of state variable would often vary on a continuous, rather than discrete, domain - e.g.~neuron membrane potential or local field potential voltages. The IIT algorithm cannot be applied when state variables are taken to be continuous, because of the use of a uniform (maximum entropy) distribution over states, which is not defined for continuous variables \citep{Barrett2019}. (Unless one can define precise hard bounds on their range, which is not the case for neuron membrane potentials or local field potentials.) In theory, the IIT algorithm (implicitly) considers only discrete choices of state variables. However, it is uncertain if quantities are guaranteed to remain bounded as the number of discrete states in the state graining is increased.
   
\subsection{Proxies are not approximations} \label{sec:proxy}
We have frequently heard and seen claims that approximations to $\Phi$ can be successfully applied to experimental data, e.g.~in \citet{Tononi2022b}. However, this use of the word \textit{approximation} is incorrect, and overstates what has been (and what can be) achieved. The word \textit{proxy} would be more accurate, and would prevent overclaiming about the degree to which empirical evidence supports the theory. For one quantity $A$ to be an approximation to another $B$, $A$ and $B$ must both be well-defined, and one must be able to determine a particular bound on the maximal possible discrepancy between $A$ and $B$. To illuminate the distinction between a proxy measure and an approximation of a measure, here are some examples of each.

In the field of economics, the gross domestic product (GDP) is often used as a \textit{proxy} measure for the overall standard of living of a country. However GDP does not directly measure the standard of living. While various measures of individual happiness and wellbeing exist, it is impractical to measure these for every individual in a country, and there is no precise formula relating the sum of any happiness or wellbeing index to GDP, even in theory. It would thus be wrong to say that GDP approximates the standard of living. Nevertheless, it is commonly used as a proxy, because it is a practical way of gaining some information about the overall state of a country.

In physics, the trajectory of a rocket fired into space can be calculated most accurately using Einstein's general theory of relativity. However, Newton's theory of gravitation can provide an \textit{approximation} to the trajectory, and it can be established how accurate this approximation is. Specific bounds can be placed on the maximum deviation the rocket will exhibit from the trajectory calculated using Newton's theory.

One measure from information theory that is frequently computed in \textit{approximation} is the transfer entropy between two variables \citep{Schreiber2000}. This has been applied extensively to neurophysiological data. Various methods are used to empirically approximate (\textit{estimate}) what would be the transfer entropy in the limit of infinite data, under assumptions of stationarity. Whichever approximation method is used, it is known that there is theoretical convergence in the limit of infinite data. Furthermore, Granger causality is often used as an approximation to transfer entropy, that specifies just the contribution from fitting a linear model to the data and neglecting non-linear components in the dynamics \citep{Barnett2015}.

At the same time, the transfer entropy between two recorded EEG time series would nevertheless only be a \textit{proxy} for the rate of transfer of information from one underlying region of cortex to another. There is no precise mathematical link at all between any component of an EEG signal and information storage or propagation in actual neurons (assuming such quantities are well-defined). 

Considering IIT, since $\Phi$ is not well-defined for the human brain (or anything closely resembling a human brain), one cannot approximately compute $\Phi$ from neuroimaging data. What has been applied are \textit{proxy} measures, that have some, but not all of the properties required by IIT for a quantity to serve as a fundamental measure of consciousness. Specifically, such measures typically take into consideration functional segregation and/or integration, and thus serve as operational proxies for at most two of the total set of six fundamental properties (axioms) specified by IIT.

The most successful proxy measure is PCI. PCI measures signal diversity in the EEG response to TMS, and there have been some proof-of-principle examples in logic-gate systems relating this to an earlier version of $\Phi$ \citep{Sevenius}. However, it is not possible to mathematically link PCI and $\Phi$ on any realistic model at the whole-brain level, because such a model would be non-Markovian and hence $\Phi$ would not be well-defined (see Section \ref{sec:undefined}). 

In addition to PCI, several `empirical' $\Phi$ proxies have been studied on some simple model dynamical systems, and their behaviour has been found to vary substantially \citep{Mediano2019}, with distinct measures often moving in different directions under the same change of parameters. Thus, caveat is needed when interpreting the empirical behaviour of any proxy measure; and validation with models is important. 

In summary, it would be useful if researchers were clear that empirical work uses proxy $\Phi$ measures, and not $\Phi$ itself, nor approximations to $\Phi$. This will create better understanding of the empirical status of IIT.


\subsection{In fundamental physics, the standard model is built from continuous fields, not discrete logic gates or neurons} \label{sec:fields}


IIT makes use of a mathematical formalism that requires discrete units updating their state in discrete time. This seems incompatible with the aim of IIT to describe the fundamental physical substrate of consciousness. In physics, the most successful fundamental theories have been cast in terms of continuous fields evolving in continuous time --- for example Einstein's theory of mass and gravitation (general relativity), Maxwell's theory of electromagnetism, and the standard model of particle physics \citep{Barrett2014, Barrett2016}. At the Planck scale, where quantum gravity is relevant, it is possible that space and time become discrete, but the effective physics across all observable scales in the universe is described at the fundamental level by field theories (and moreover, IIT makes no claim to justify its discrete formal basis in terms of a quantum-level physical reality).

In fundamental physics, there are fields associated with each of the fundamental particles (particles can be conceived as waves that propagate across fields). One can posit that all fundamental fields have the potential to generate consciousness, and then observe that the electromagnetic field is unique in that its physics enables complex configurations to arise \citep{Barrett2014}. It has therefore been suggested to re-formulate IIT to produce formulae for mapping configurations of fields to qualia \citep{Barrett2014}. Importantly, such a theory would not do away with neurons as playing a crucial role in generating consciousness. In this view, neurons are the scaffolding that enable a complex electromagnetic field to arise in the brain. However, it is the electromagnetic field that is the fundamental physical entity, and not the neuron. A reformulation of IIT in terms of fields could potentially lead to improved proxy measures for application to data, since observable variables in neuroimaging experiments are typically best modelled as continuous rather than discrete (see Section \ref{sec:notcomputed}).

\section{Discussion}

IIT takes an ambitious and rigorous approach to explaining consciousness. However, its ambition is also the source of the main challenges it faces.

IIT purports to be a fully comprehensive description of the physical substrate of consciousness. However, in its current form it is not compatible with fundamental physics (specifically, to the fundamental fields) and some of its key mathematical quantities are not well defined for real physical systems (such as brains).

It has been claimed that IIT enjoys empirical validation. There are indeed results that are encouraging for IIT. However, the empirically applicable measures inspired by IIT, such as PCI, are proxies rather than approximations, limiting their interpretive value. In addition, the empirical evidence gathered so far is compatible with both weak and strong versions of IIT (and potentially with other theories of consciousness as well). 

Nevertheless, even without a well-defined mathematical formalism, IIT has successfully built potential links between neural architectures and pervasive aspects of contents of consciousness, specifically space and time, with ongoing work targeting potential architectures for further aspects such as conceptual hierarchy and narrow qualia (e.g.~colour). In addition, the form of panpsychism implied by IIT is perfectly compatible with empirical science. Finally, contrary to much commentary, IIT does not attribute substantial consciousness to an inactive grid of logic gates of any size.

Arising from our exploration of the good, the bad and the misunderstood in IIT are several recommendations which may be helpful for increasing understanding of IIT and further development of the theory. These are summarised in Box 2. 

We are agnostic on whether it will be possible to revise the IIT algorithm to render it both mathematically well-defined and compatible with current models of fundamental physics. If such a revision were possible, for credence, there would then be the challenge to prove that it is the only such algorithm consistent with its axioms \citep{Barrett2019}.

We have not contested the axioms, or indeed the axiomatic approach in this paper. For a discussion on potential problems with the axiomatic approach, see \citet{Bayne2018}. We have noticed that sometimes when researchers question whether certain axioms really are self-evident and universal properties of conscious experience, there can be some misinterpretation. To address this, where we stated the axioms in Box 1, we added in some details. Arguably, the only axiom that is not obvious from introspection is \textit{exclusion}, although this axiom remains a reasonable assumption. (\textit{Exclusion} implies that whether or not a system is a substrate of consciousness is determined by matter external to it, and \citet{Morch2019} argued that this is inconsistent with the intrinsicality axiom. However, it can be argued that \textit{intrinsicality} does not imply that consciousness can't be impacted by external matter; but only that consciousness is independent of any point of view, i.e.~frame of reference, of any external observer. On this understanding of \textit{intrinsicality}, there is no contradiction between these two axioms.) We have also not discussed how additional axioms may help the theory. See \citet{Singhal2022} for a discussion on how additional axioms might be needed for IIT to account for all components of temporal phenomenology. In particular, it is noted there that, in isolating temporal grains of experience to a single timescale, IIT misses out on capturing the multi-scale nested nature of temporal phenomenology.

 



\begin{boxA} \label{box:recommendations}
\subsubsection*{Box 2. Summary of recommendations for future development of IIT.}
\begin{itemize}
\item \textbf{Multi-dimensional measures of the state of consciousness.} To avoid further misunderstandings around the quantity of consciousness, the structure integrated information $\Phi$ might be replaced with a suite of quantities, e.g. (i) mean $\Phi$ over a period of time; (ii) variances over time of some quantities within the cause-effect structure; (iii) some measure of (mean) complexity of the geometry and topology of the cause-effect structures. Doing this would potentially enable the theory to better characterise the richness of consciousness for any physical substrate that endures in time, such as the human brain. See Section \ref{sec:highphi}.
\item \textbf{Communicate clearly about the panpsychism implied by IIT.} Critiques of IIT for being panpsychist ought to be avoidable if the precise flavour of panpsychism implied by the theory is clearly communicated. See Section \ref{sec:panpsychism}.
\item \textbf{Reformulate the algorithm so it involves only the current state.} This would constitute a major change to the theory, but is possibly the only way to make the algorithm well-defined for all (non-Markovian) systems. See Section \ref{sec:undefined2}.
\item \textbf{Emphasise that empirical work uses proxy $\Phi$ measures.} To create better understanding of the empirical status of IIT, the word approximation should be avoided unless fully justified. See Section \ref{sec:proxy}.
\item \textbf{Reformulate the theory to apply to fields.} For compatibility with the standard model of particle physics, the theory might be revised so all quantities are computed based on the configuration of the electromagnetic field of the system of interest, as opposed to discrete states of discrete system components. See Section \ref{sec:fields}.
\end{itemize}
\end{boxA}

\vspace{0.3cm}

Some of us have previously suggested a distinction between two flavours of IIT: strong IIT, which seeks a \textit{structural identity} between consciousness and physical structures based on `intrinsic' information; and weak IIT, which more modestly seeks out \textit{explanatory correlations} between various aspects of consciousness and broader measures of information dynamics with explanatory and predictive value \citep{Mediano2022}. Strong IIT is appealing for those who prefer to tackle the problem of the fundamental physical nature of consciousness with theoretical methods from mathematics, physics and philosophy. Future work might attempt a formulation in terms of fields to link the theory to the current standard model of particle physics \citep{Barrett2014}. Having a conceptually related, yet distinct, weak IIT will allow researchers to work on developing and experimentally testing various intuitions of IIT without necessarily committing to the more radical propositions of strong IIT (notably its identity claim) and/or having to address its open mathematical problems.  This might help catalyse a wider uptake of IIT-related ideas in the community.

It is important to note that the recent adversarial experiments run by the COGITATE consortium were only testing some of the intuitions of IIT, such as where in the brain the physical substrate of consciousness is located and which patterns of functional connectivity should be associated with consciousness, and did not make use of any proxy measures of integrated information \citep{Melloni2025}. The INTREPID consortium are seeking to test whether conscious contents differ when certain neurons have been inactivated compared to when they are simply inactive \citep{Olcese2024} - if this is successful, this will constitute an empirical test of a core claim of IIT, but this again does not rely on the mathematical measures in strong IIT being well-defined. It is important to note here that empirical deactivation of neurons is also likely to be an approximation, since baseline neural activity (inactive, but not inactivated) is always more than zero.


IIT is unique in how it tries to derive a formula, from phenomenology,  for determining and describing the fundamental physical substrate of consciousness. While there are several problems with it, there are also enough successes to show that the approach has merit, providing inspiration for future developments of the theory, or for successor theories  \citep{Cleeremans2025}. At the same time, it is important to remember that even if (some future version of) IIT were all correct, there may be limits to its reach. While IIT explains certain fundamental aspects of phenomenology in terms of physical mechanisms and information dynamics, there are several important properties of consciousness in humans and other animals, such as metacognition, sense of self, and capacity for suffering, which so far have not been addressed. We hope the arguments in this article increase understanding around IIT, that our recommendations help IIT-inspired research evolve fruitfully, and that together they provide a useful reminder of the rigorous scientific research and debate that IIT has inspired.

\section*{Acknowledgements}
AKS is supported by ERC Advanced Investigator Grant 101019254, under the Horizon 2020 research programme.


\section*{Data availability}
There are no new data associated with this article.


\begin{thebibliography}{99}

\bibitem[Aaronson(2014)]{Aaronson}
Aaronson, S. (2014) Why I am not an integrated information theorist (or, the unconscious expander). \textit{Shtetl-Optimized}. https://scottaaronson.blog/?p=1799

\bibitem[Albantakis et al.(2023)]{Albantakis2023}
Albantakis, L., Barbosa, L., Findlay, G., Grasso, M., Haun, A.M., Marshall, W., et al. (2023) Integrated information theory (IIT) 4.0: formulating the properties of phenomenal existence in physical terms. \textit{PLoS Comput. Biol.} 19(10): e1011465.

\bibitem[Barrett(2014)]{Barrett2014}
Barrett, A.B. (2014) An integration of integrated information theory with fundamental physics. \textit{Front. Psychol.} 5, 63.

\bibitem[Barrett(2016)]{Barrett2016}
Barrett, A.B. (2016) A comment on Tononi \& Koch (2015) `Consciousness: here, there and everywhere?'. \textit{Phil. Trans. R. Soc. B} 20140198.

\bibitem[Barrett and Mediano(2019)]{Barrett2019}
Barrett, A.B. and Mediano, P.A.M. (2019) The phi measure of integrated information is not well-defined for general physical systems. \textit{J. Conscious. Stud.} 26(1--2), 11--20.

\bibitem[Bartlett(2022)]{Bartlett2022}
Bartlett, G. (2022) Does integrated information theory make testable predictions about the role of silent neurons in consciousness? \textit{Neurosci. Conscious.} 2022(1), niac015.

\bibitem[Bayne et al.(2016)]{Bayne2016}
Bayne, T., Hohwy, J., and Owen, A.M. (2016) Are there levels of consciousness? \textit{Trends Cogn. Sci.} 20(6), 405--413.

\bibitem[Bayne(2018)]{Bayne2018}
Bayne, T. (2018) On the axiomatic foundations of the integrated information theory of consciousness. \textit{Neurosci. Conscious.} 2018(1), niy007.

\bibitem[Casali et al.(2013)]{Casali2013}
Casali, A.G. et al. (2013) A theoretically based index of consciousness independent of sensory processing and behavior. \textit{Sci. Transl. Med.} 5, 198ra105.

\bibitem[Casarotto et al.(2016)]{Casarotto2016}
Casarotto, S. et al. (2016) Stratification of unresponsive patients by an independently validated index of brain complexity. \textit{Ann. Neurol.} 80, 718--729.

\bibitem[Cea et al.(2023)]{Cea2023}
Cea, I., Negro, N., and Signorelli, C.M. (2023) The fundamental tension in integrated information theory 4.0's realist idealism. \textit{Entropy} 25, 1453.

\bibitem[Chalmers(1995)]{Chalmers1995}
Chalmers, D.J. (1995) Facing up to the problem of consciousness. \textit{J. Conscious. Stud.} 2, 200--219.

\bibitem[Cleeremans et al.(2025)]{Cleeremans2025}
Cleeremans, A., Mudrik, L., and Seth, A.K. (2025) Consciousness science: where are we, where are we going, and what if we get there? \textit{Front. Sci.} 3. doi:10.3389/fsci.2025.1546279.

\bibitem[Cogitate Consortium et al.(2025)]{Melloni2025}
Cogitate Consortium., Ferrante, O., Gorska-Klimowska, U. et al. (2025) Adversarial testing of global neuronal workspace and integrated information theories of consciousness. \textit{Nature}. https://doi.org/10.1038/s41586-025-08888-1.


\bibitem[Comolatti et al.(2024)]{Comolatti2024}
Comolatti, R., Grasso, M., and Tononi, G. (2024) Why does time feel the way it does? Towards a principled account of temporal experience. \textit{iScience} 28, 113434.

\bibitem[Farisco and Changeux(2023)]{Farisco2023}
Farisco, M. and Changeux, J.P. (2023) About the compatibility between the perturbational complexity index and the global neuronal workspace theory of consciousness. \textit{Neurosci. Conscious.} 2023(1), niad016.

\bibitem[Ferrarelli et al.(2010)]{Ferrarelli2010}
Ferrarelli, F. et al. (2010) Breakdown in cortical effective connectivity during midazolam-induced loss of consciousness. \textit{Proc. Natl. Acad. Sci.} 107, 2681--2686.

\bibitem[Findlay et al.(2024)]{findlay2024dissociating}
Findlay, G., Marshall, W., Albantakis, L., David, I., Mayner, W.G.P., Koch, C., and Tononi, G. (2024) Dissociating artificial intelligence from artificial consciousness. arXiv:2412.04571.

\bibitem[Fuli{\'n}ski et al.(1998)]{Fulinski:1998}
Fuli{\'n}ski, A., Grzywna, Z., Mellor, I., Siwy, Z., and Usherwood, P.N.R. (1998) Non-Markovian character of ionic current fluctuations in membrane channels. \textit{Phys. Rev. E} 58(1), 919.

\bibitem[Gomez-Marin and Seth(2025)]{Gomez-Marin2025}
Gomez-Marin, A. and Seth, A.K. (2025) A science of consciousness beyond pseudo-science and pseudo-consciousness. \textit{Nat. Neurosci.} 28, 703--706.

\bibitem[Haun and Tononi(2019)]{Haun2019}
Haun, A. and Tononi, G. (2019) Why does space feel the way it does? Towards a principled account of spatial experience. \textit{Entropy} 21(12), 1160.

\bibitem[IIT-Concerned et al.(2025)]{Hakwan}
IIT-Concerned., Klincewicz, M., Cheng, T. et al. (2025) What makes a theory of consciousness unscientific? \textit{Nat. Neurosci.} 28, 689--693.


\bibitem[Kleiner and Tull(2021)]{Kleiner21}
Kleiner, J. and Tull, S. (2021) The mathematical structure of integrated information theory. \textit{Front. Appl. Math. Stat.} 6, 2020.

\bibitem[Kuhn(2024)]{Kuhn2024}
Kuhn, R.L. (2024) A landscape of consciousness: toward a taxonomy of explanations and implications. \textit{Prog. Biophys. Mol. Biol.} 190, 28--169.

\bibitem[Levine(1983)]{Levine1983}
Levine, J. (1983) Materialism and qualia: the explanatory gap. \textit{Pac. Philos. Q.} 64, 354--361.

\bibitem[Marshall et al.(2024)]{Marshall2024}
Marshall, W., Findlay, G., Albantakis, L., and Tononi, G. (2024) Intrinsic units: identifying a system's causal grain. \textit{bioRxiv} 2024.04.12.589163.

\bibitem[Massimini et al.(2005)]{Massimini2005}
Massimini, M. et al. (2005) Breakdown of cortical effective connectivity during sleep. \textit{Science} 309, 2228--2232.

\bibitem[Massimini et al.(2009)]{Massimini2009}
Massimini, M. et al. (2009) A perturbational approach for evaluating the brain's capacity for consciousness. \textit{Prog. Brain Res.} 177, 201--214.

\bibitem[Mediano et al.(2019)]{Mediano2019}
Mediano, P.A.M., Seth, A.K., and Barrett, A.B. (2019) Measuring integrated information: comparison of candidate measures in theory and simulation. \textit{Entropy} 21, 17.

\bibitem[Mediano et al.(2022)]{Mediano2022}
Mediano, P.A.M., Rosas, F.E., Bor, D., Seth, A.K., and Barrett, A.B. (2022) The strength of weak integrated information theory. \textit{Trends Cogn. Sci.} 26(8).

\bibitem[M{\o}rch(2019)]{Morch2019}
M{\o}rch, H.H. (2019) Is consciousness intrinsic? A problem for the integrated information theory. \textit{J. Conscious. Stud.} 26(1--2), 133--162.

\bibitem[Mudrik et al.(2025)]{Mudrik2025}
Mudrik, L. et al. (2025) Unpacking the complexities of consciousness: theories and reflections. \textit{Neurosci. Biobehav. Rev.} 170, 106053.


\bibitem[Olcese et al.(2024)]{Olcese2024}
Olcese, U., Muckli, L.F., Boly, M. et al. (2024) Accelerating research on consciousness: an adversarial collaboration to test contrasting predictions of the integrated information theory and predictive processing accounts of consciousness -- version 2. Open Science Foundation. https://doi.org/10.17605/OSF.IO/4RN85.

\bibitem[Re et al.(2021)]{Re2021}
Re, V.L. et al. (2021) Role of transcranial magnetic stimulation (TMS) combined with electroencephalography (EEG) in disorders of consciousness (DOC). \textit{J. Neurol. Sci.} 429, 118507.

\bibitem[Sarasso et al.(2021)]{Sarasso2021}
Sarasso, S., Casali, A.G., Casarotto, S., Rosanova, M., Sinigaglia, C., and Massimini, M. (2021) Consciousness and complexity: a consilience of evidence. \textit{Neurosci. Conscious.} 2021(2), niab023.

\bibitem[Schreiber(2000)]{Schreiber2000}
Schreiber, T. (2000) Measuring information transfer. \textit{Phys. Rev. Lett.} 85(2), 461--464.

\bibitem[Seth(2009)]{Seth2009}
Seth, A.K. (2009) Explanatory correlates of consciousness: theoretical and computational challenges. \textit{Cogn. Comput.} 1, 50--63.

\bibitem[Seth et al.(2015)]{Barnett2015}
Seth, A.K., Barrett, A.B., and Barnett, L. (2015) Granger causality analysis in neuroscience and neuroimaging. \textit{J. Neurosci.} 35(8), 3293--3297.

\bibitem[Seth and Bayne(2022)]{Seth2022}
Seth, A.K. and Bayne, T. (2022) Theories of consciousness. \textit{Nat. Rev. Neurosci.} 23, 439--452.

\bibitem[Sevenius Nilsen et al.(2019)]{Sevenius}
Sevenius Nilsen, A., Juel, B.E., and Marshall, W. (2019) Evaluating approximations and heuristic measures of integrated information. \textit{Entropy} 21(5), 525.

\bibitem[Singhal et al.(2022)]{Singhal2022}
Singhal, I., Mudumba, R., and Srinivasan, N. (2022) In search of lost time: integrated information theory needs constraints from temporal phenomenology. \textit{Philos. Mind Sci.} 3. doi:10.33735/phimisci.2022.9438.

\bibitem[Tononi and Edelman(1998)]{Tononi1998}
Tononi, G. and Edelman, G.M. (1998) Consciousness and complexity. \textit{Science} 282(5395), 1846--1851.

\bibitem[Tononi(2004)]{Tononi2004}
Tononi, G. (2004) An information integration theory of consciousness. \textit{BMC Neurosci.} 5, 42.

\bibitem[Tononi et al.(2016)]{Tononi2016}
Tononi, G., Boly, M., Massimini, M. et al. (2016) Integrated information theory: from consciousness to its physical substrate. \textit{Nat. Rev. Neurosci.} 17, 450--461.

\bibitem[Tononi et al.(2022)]{Tononi2022}
Tononi, G., Albantakis, L., Boly, M., Cirelli, C., and Koch, C. (2022) Only what exists can cause: an intrinsic view of free will. arXiv:2206.02069.

\bibitem[Tononi et al.(2022b)]{Tononi2022b}
Tononi, G., Boly, M., Grasso, M. et al. (2022b) IIT, half masked and half disfigured. \textit{Behav. Brain Sci.} 45, e60.

\bibitem[Tononi et al.(2025)]{Tononi2025}
Tononi, G., Albantakis, L., Barbosa, L. et al. (2025) Consciousness or pseudo-consciousness? A clash of two paradigms. \textit{Nat. Neurosci.} 28, 694--702.

\end{thebibliography}
\end{document}